# PHISHING MITIGATION TECHNIQUES: A LITERATURE SURVEY

Wosah Peace Nmachi and Thomas Win

School of Computing & Engineering University of Gloucestershire, Park Campus, Cheltenham GL50 2RH United Kingdom

## ABSTRACT

*Email is a channel of communication which is considered to be a confidential medium of communication for exchange of information among individuals and organisations. The confidentiality consideration about e-mail is no longer the case as attackers send malicious emails to users to deceive them into disclosing their private personal information such as username, password, and bank card details, etc. In search of a solution to combat phishing cybercrime attacks, different approaches have been developed. However, the traditional exiting solutions have been limited in assisting email users to identify phishing emails from legitimate ones. This paper reveals the different email and website phishing solutions in phishing attack detection. It first provides a literature analysis of different existing phishing mitigation approaches. It then provides a discussion on the limitations of the techniques, before concluding with an explorationin to how phishing detection can be improved.*

## 1. INTRODUCTION

Phishing is one of the specific types of social engineering attacks that are well known globally for bypassing deploy technical defenses by manipulating object characteristics such as system applications or platforms to deceive, rather than directly attack the targeted user (Ryan & George, 2015). It is common among other security threats, use as the initial step to gain access to an electronic device for further exploitation without the user awareness. Phishers deceive people and obtain secret information [1], such as usernames, passwords, credit card numbers, and IDs from a victim[36]. It targets the human element of cyber-securitywhich[37]account for 95% of cyber incidents and is used as the initial stages usedin cyber-security breaches [38],[21],[39].According to the UK Cyber-security and Strategy 2016-2021 and world statistics, almost all the successful cyber-attacks have a contributing human influence [40] which is to say that cyber-security is not just about the technology as human knowledge on security is also required for cyber-security stability. When an email gets to a user email-box, it is the user that reads and responds to it and where a malicious email is ignored by the user the attack is killed instantly and no loss. The security of cyber environment is not stable as attackers are messing the environment up at will making the goal of cyber-security look like it is unachievable. There have been different countermeasures which have been proposed to mitigate phishing attacks. However, these solutions have not achieved the expected decrease of phishing attacks due to the fact that the human security factors that phishers exploit often have not received an easy to use and identify phishing emails among genuine ones[17]. Human contributions in form of knowledge will go a long way in curbing cyber-attacks asknowledgeis said to be power. Therefore, thetrainingusers approach has been adopted by many organisations and research [16], [42] with the aim of improving the human knowledge on cyber-security through raising awareness. However, retention of the knowledge gained is a challenge to this approach as users seem to forget some of the knowledge and information related to security awareness [21][22]. Therefore, email users need to be assisted in identifying phishing emails. This paper provides a literature review of the different approaches and techniques which are proposed in existing research in phishing detection.

This paper is organized as follows: Section 2talks about thestate of the art of the different phishing detection approaches in existing research. Section 3 discusses the strengths and limitations of each approach in this research and Section 4concludes the paper.

## 2. STATE OF THE ART

The increase in phishing attacks attracted much attention of researchers to this area of interest. To mitigate phishing attacks the existing approaches can be categorised into four groups namely: stylometric analysis, rule-based, classification-based, and user education. Figure 1 provides a graphical overview of how a topical phishing attack operates.

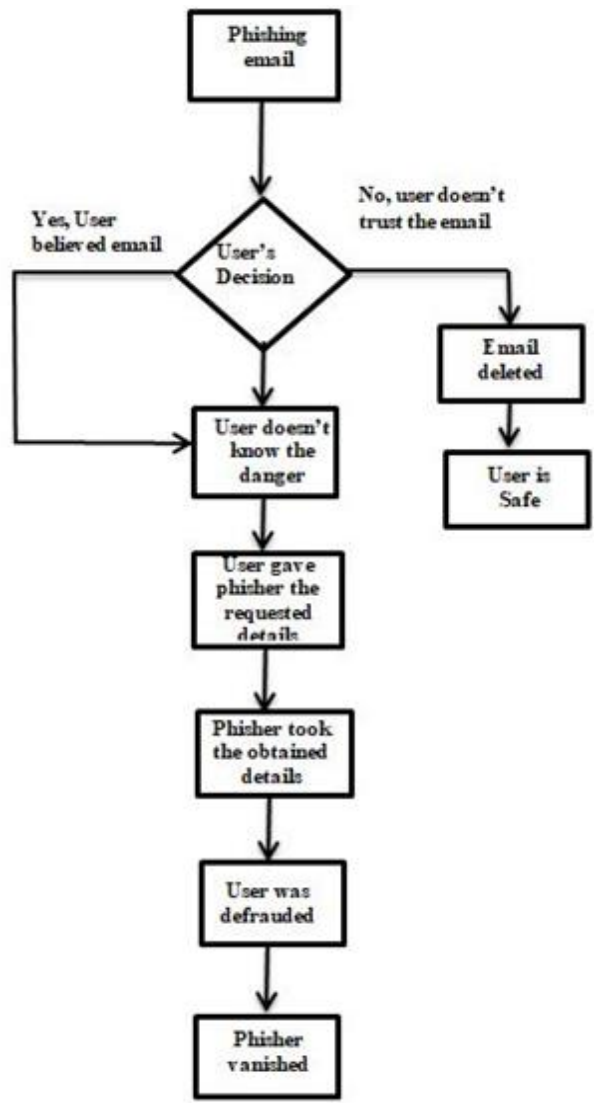

Figure 1.Overview of a phishing attack

## 2.1. Stylometric Analysis

The stylometric analysis involves analyzing the unique writing behaviour and linguistic styles of users to ascertain authorship. It assumes that an individual author displays a specific writing habit that will be captured as features, such as phraseology, core vocabulary usage, and sentence complexity [4]. In[6] their work presented a model to detect spear-phishing emails which the authors sent to employees of 14 international organisations, by using a combination of stylometric features extracted from email subjects, bodies, and attachments, andsocial features from LinkedIn profiles and according to the evaluated results, both achieved an overall accuracy of 97.76% in detecting spear-phishing emails. However, email features achieved a slightly better accuracy of 98.28% without the social features. It was noted that the features obtained from LinkedIn did not help in identifying spear-phishing emails. The approach proposed in [22]clearlyinformed the users of possible mismatches between the writing styles of a received email body and of trusted authors by going through the email received and study the email body itself to capture the writing style of the sender. The experimental implementation was conducted using the source-code authorship technique, called Source Code Author Profiles (SCAP), and the dataset used for the assessment was created from email messages extracted from 12 authors' email inbox which amounted to 289 emails and it gave an accuracy of 87% of authorship prediction accuracy of email messages. This SCAP method used in this framework gave a high false negative rate as it is originally designed for software source codes authorship. Hence, the authors suggested the use of alternative methods as no mandated specific method is tied to the implementation of this work. The framework complemented the user ID-based authentication techniques and further enhanced the security in an easy to use manner. In[23] their work mined the writing styles of email users from a collection of e-mails written by multiple unknown authors. The whole idea behind their work is to first cluster the anonymous e-mails by the stylometric features and then extract the writing style of authors from each cluster. They argued that the presented problem together with their proposed solution is different from the traditional problem of authorship, which assumes training data is always available for building classifiers. The proposed technique specifically helps out in the initial stage of investigating any case involving anonymous emails, in which the investigator has little information on the case and the authors of the suspicious email. The experiment conducted on a real dataset suggested that clustering by writing style is a promising approach for grouping emails authored by the same user.

In [3]a novel automated approach to defend users against spear-phishing attacksispresented. This approach involves building probabilistic models of both email metadata and stylometric features of email content using natural language processing. Subsequent emails are compared to these models which are developed using a Support Vector Machine for classification. [2]improved it by combining stylometric features, gender features, and personality features.Their approach uses feature extraction to build and keep an identity profile model of a sender; hence subsequent emails of the sender are compared against the profile, and in a case where the profile of an uncertain email is consistent with the legitimate profile of the sender, the sender of the uncertain email is identified as legitimate and the email is considered normal mail. However, if the profile of an uncertain email is inconsistent with the legitimate profile of the sender, the sender is masked, and the email is classified asaspear-phishing email. The experimental result showed the detection accuracy of 95.05% using the Enron email dataset which was gathered through theaCognitive Assistant that Learns and Organizes (CALO) project.The stylometric analysis focuses on the email sender's writing pattern and the mail sender is discriminated based on the similarity of mails characterized with stylometric features.

## 2.2. Rule-Based Approaches

Rule-basedsolutions include Blacklist and Whitelist technologies: both blacklist and whitelist are used to prevent phishing attacks by keeping a dataset of trusted and untrusted websites[19], and email addresses. Blacklists block content based on pre-defined malicious IP address, Universal Resource Allocator (URL), email address and a few keywords, as well as user behaviours such as click, update, provide, follow, link, etc. With blacklist, every requested URL and sender addressesare being compared to a listed phishing blacklist and where an internet user tries to visit a fake website which is already known or exiting in the blacklist, the web browser stops the user from visiting by denying access, and unsolicited emails are blocked by blacklist based on fake sender addresses.Whitelists on the other hand only keep list of trusted websites and user can only be allowed to access a website which is approved to be a legitimate site[19].Whitelistsrecognise trusted sender email addresses as well. Both whitelists and blacklists perform well with a well-known or identified phishing website, etc. A trusted website which is not listed in the whitelist and a user tries to access it the whitelistwill consider it to be a phishing site because it is not known by the whitelist making the false negative rate of this approach to be very high. DNS-based blacklist, this approach blacklists a range of IP addresses and domain names [5]. Google Safe Browsing API provides a blacklist and browser blocks page if there are any hits [12].PhishNet exploits the observation that attackers frequently employ simple modifications, changing top level domain to URLs. The authors in their work proposed five heuristics to enumerate simple combinations of known phishing websites to discover new phishing URLs and applied a matching algorithm to analyze a URL by taking it apart into multiple components that are matched individually against all entries in the blacklist [44]. Spoof Guard detects spoofed pages based on URLs with the help of set rules [43]. Domain-level authentication is utilised by sending Domain Key, and for it to work both sides, sender and receiver must use the same technology[45]. Sender ID is implemented on Microsoft sender ID and it works when both sides have the same technology. It is used at domain level authentication for sending Domain Key [24]. PhishGuard: with this, phishing websites do not respond correctly while requesting credentials.Phish-wish: it is a stateless phishing channel using negligible principle, it has low false positive [25]. The rule-based approaches are among the earliest solutions which are proposed for spam detection [47].This approach performs well on known set rules. However, it has high false alarm rates and its difficulty in updating rules in case of big data is also a challenge to this approach [2].Therefore, a research in the area of rule-based to reduce the time it takes to update rules would definitely help in combat phishing attack as black and whitelists have proven to be effective approaches to phishing attacks but the manual updating did not help it, hence a need for automatic rule update system in the organizations for the safety of the users.

## 2.3. Classification-Based Approaches

Classification-based solutions involve using machine learning techniques such as classification or clustering for phishing detection. A classification approach featuring Support Vector Machine(SVM) is used to develop a classification model using structural properties found in Mail Transfer Agent (MTA) and Mail User Agent(MUA) [20][7].Theproposed approach intercepts each ongoing email and checks for any phishing attribute and characteristic using the trained SVM classifier. Similarly [18]also adopted the idea of using an SVM to read the email messages and explore the email attributes and characteristics furthermore to detect spear-phishing emails. K-Nearest Neighbour (KNN)is used to rank emails as either spam, ham, or phishing. It detects emails based on similarities in k-sample phishing emails[5]. Decision Tree Algorithm (DTA) is implemented to detect malicious attachment files of phishing websites in the email bodies to prevent the user from falling for such attack [8]. To detect phishing attacks, [19] combined the reinforcement learning with neural network approach for the classification of

emails. In [26]Natural Language Processing (NLP) is usedto detect phishing emails. The authors focused their approach on the natural language text in the attack to perform a semantic analysis of the email textin order to detect malicious intend. [27], proposed a real-time phishing detection system, which uses seven different classification algorithms, such as, Random Forest, Naive Bayes, Adaboost, k-Nearest Neighbours (kNN), K-star, Decision Tree, and Sequential Minimal Optimisation (SMO)andNatural Language Processing (NLP) based features.[46], proposed a semantic-based classification approach to improve the spam detection accuracy. The authors composed this approach in two stages where the first stage classifies contents of email by subject domains, and the second stage builds domain-specific semantic features on which spam classification is carried out. The authors used and compared different machine learning classification algorithms and the best classifier giving the most precise email classification is identified. The experimental results of this approach have shown to outperformed several existing methods based on Bag of Word (BoW) and latent semantic analysis.

CANTINA Searches top Term Frequency-Inverse Document Frequency (TF-IDF) in Search Engine (SE) like Goggleand finds current URLs in the top list. It has a high false positive rate when Term Frequency (TF) of any other term is high [28] as only the top N-terms with highest values are used to represent any document. Visual similaritycomparison is considered to be an effective anti-phishing approach for phishing attack detection by comparing the visual appearance and similarity between the spoofed site andaphishing webpage using, images embedded in the page, features-text pieces, and visual appearance of the page are considered for comparing the similarity. However, it compares only content on the websites and has a high false positive rate [29]. In [33]an anti-phishing approach is proposed which uses deep semantic analysis, and used both machine learning, and deep learning techniques, to capture inherent users' email texts and classify them as either phishing or legitimate email. The result of their work shows that deep learning models performed a little better than the machine learning models.

Apart fromtheSupport Vector Machine which the accuracy was lightly better with word phrasing than without word phrasing. It was found that the context of the email language is important in identifying phishing emails from legitimate ones.Jain and Gupta [34] proposed a client-side and no third-party services required approach to detect phishing attacks by analysing the hyperlinks found in the HTML source code of the website. It is language-independent and can detect websites written in any textual language. The approach used hyperlink features and grouped the features into 12 different categories and used the same features to train the machine algorithms. The methodisevaluatedonvarious classification algorithms using legitimate and non-legitimate websites dataset to see which classifier achieves a better result and logistic regression classifier was stated to have achieved a better accuracy of 98.42% which is more than other classifiers in the detection of phishing websites.In[35]a deep-spam-phish-net, a framework for phishing and spam detectionisproposed. The framework has two sub-modules. The first sub-module detects phishing and spam emails and the second sub-module detects phishing and spam URLs using various deep learning architectures for both Phishing and Spam detection with emails and URL data sources. They used various datasets collected from both public and private data sources and the datasets were used for experiments with deep learning architectures. All the experiments conducted ran to 1,000 epochs with different learning rates ranging from 0.01 to 0.5.

Classical machine learning algorithms and deep learning architectures are compared in the conversion of text data into numeric representation and various natural language processing text representation methods were used. The performances of machine learning and deep learning architectures algorithms were evaluated in each module and in most of the cases it showed that the deep learning architectures outperformed the machine learning algorithms when compared. However, this work was focused on phishing and spam emails and URLs detections in cyber-security. Machine learning technique can effectively detect phishing emails by deleting or adding the features extracted from the email. Attackers utilise email services like Yahoo, Gmail, etc. to achieve communication with malware using Domain Generation Algorithms to generate new domain names [31]. However, deep learning architectures are being applied for the Domain Generation Algorithm (DGA) detection to mitigate the attack[32]. Although some of the machine learning types require a large amount of dataset and computational power for better performance, such as deep learning architectures. Knowing that deep learning can learn from a vast amount of data is a good thing and it can be used to tackle the issues of phishing attacks just as some recent researches have used it and when compared with traditional machine learning, deep learning actually outperformed the traditional machine. Seeing the way phishing attack is affecting organisation security by stealing information from them and downloading malware into their system, and making online customers to have less trust on e-commerce, deep learning is a sure way out of this problem. And with the fact that businesses are moving to virtual world because of the pandemic, the need to protect internet users' security is paramount.

## 2.4. User Education

Alongside technical solutions, user education is one of the approaches used to combat cyber threats such as phishing attacks [9] by improving users' ability to detect phishing attempts [10] in order to avoid it when the trained users get confronted by either phishing emails or URLs. According to[16] ordinary web browsing users are not aware of how phishing attacks start or how to visually recognize illegitimate webpage from legitimate ones. In [13] a phishing detection application called NoPhish to detect phishing URLs is proposed. It is an application, where the users can lose or win points and the result showed promising at the time of the research. However, it is only useful to the users and knowledge retention is also a challenge to this approach. Therefore, the researchers in[11]developedHuman-as-a-Security-Sensor (HaaSS) which uses the ability of human-users as sensors that can detect and report information (security threats) accordingly and the users' reports are encouraged and taking into account to strengthen organisation cyber-security awareness. The user education technique helps internet users to be aware of the circumstances about phishing attacks, that they may be able to minimise or avoid this risk, perhaps stop it as early as possible [16]. However, it is found that users' knowledge retention is a general challenge to the user-educationapproaches[21], and the internet users need to be up to date about the new kinds of attackandalso need to read a significant amount [19]of educative security information to be safe online. In addition, there is high monetary cost demand with this approach [41]. Therefore, a solution that can mitigate phishing attacks without internet users' intervention is what organizations need to keep their customers secured in the internet space.

Existing literature works have discussed humans' inability to interact with the systems to be one of the major reasons why people still fall for phishing attacks [14],[15], and the existing solutions have not achieved the expected decrease of phishing attacks because the human security factors that phishers exploit often have not received an easy to use and identify phishing email[17]. Also, a phishing website is known forits short life span. It lasts normally about two days by leveraging DGA (Domain Generation Algorithms), which makes the phisher to disappear immediately the fraud is committed, and because of this reason, law enforcement finds it difficult to achieve its aim [16]. According to [17], email users are not well assisted by email clients in identifying phishing emails and advised that email clients should consider using feedback mechanisms to present security-related aspects to the users to make them aware of the characteristics of phishing attacks. The email users however need all the assistance possible from their email clients to avoid phishing attack because when a phishing attack occurs, all the technical protection systems deployed cannot stop a user from disclosing requested information to the phisher over the phone or email.

## 3. CRITICAL EVALUATION

Table 1. Summary of existing approaches and their strengths/limitations

| Approaches | Strengths | Limitations |
|---|---|---|
| Stylometric Analysis | -Reveals identity<br>-Useful for spear phishing and whaling phishing attacks detection | -Change in writing could cause misclassification<br>-Small email size affects accuracy |
| Rule-Based Approach | -Performs well on known set rules<br>-Easy to manage | -It has high false alarm rate<br><br>-Difficulty in updating rules |
| Classification-Based Approach | -Can effectively detect phishing emails<br>-Can catch newly created phishing URLs | -It requires a large amount of data and high computational power |
| User Education | -Improves users' ability to detect phishing emails<br>-It educates novice users about phishing attacks | -Lack of knowledge retention<br><br>-It attracts expense |

StylometricAnalysisis an identity revealer as it involves analysing the unique writing behaviours and linguistic styles of users to ascertain authorship. It assumes that an individual author displays a specific writing habit that will be captured as features, such as phraseology, core vocabulary usage and sentence complexity [4].Spear-phishing and whalingare more akin to impersonation and identity hiding type of cyber-attacks which makes it harder to identify by users when trapped in this kind of attack. In order to reveal the impersonators' true identity, the use of stylometry approach is a good step to unmask the attackersby displaying their writing behavioursimmediately without delay. Therefore, the use of stylometric to fight phishing attacks is very much in order[3]. However, the approach has only been used much in authorship identifications. The stylometric analysis approach in email focuses on the email sender's writing pattern and the mail sender is discriminated based on the similarity of mails characterised with stylometric features. Therefore, where a user's writing pattern changed it could cause misclassification of mail.An intelligent system that can identify a user'sdifferent written styles would go a long way toward mitigating phishing attacks. Therefore, more research is highly required in this area and availability of big email datasets should also beconsidered for the use of better classifiers for better results

Rule-based approacheswork based on set rules and the approach mainly focused on blacklist and heuristic-bases. Blacklists always compare a requested URL to the existing URLs in the lists, and where there is a match the browser sends a warning to the user not to consent to the request, and if there isno match, blacklists consider it genuine even when it is harmful because blacklists are limited in deleting newly created phishing websites. The limitations of blacklists brought about the heuristic-based approach andthe heuristic-based approaches came with the ability to recognize newly phishing websites which blacklist is not able to do[30], and the ability of the heuristic-based improved the rule-based method of combating phishing attacks. In general, the rule-based approaches perform well on known set rules. However, it has high false alarm rates and its difficulty in updating rules in case of big data is also a challenge to this approach [2]. The updating issue with rule-based approaches if solved would reduce if not eliminated the false-alarm rate of this approach and the performance accuracy would as well increase. Therefore, more research in this area using Deep Learning (DL) should be considered to improve this approach and enhance cyber-security.

The most now use approach is classification-based approach. It involves using machine learning techniques such as classification or clustering for phishing detection. The machine learning techniques can effectively detect phishing emails by deleting or adding the features extracted from the email. However, the features are manually selected which is one of the limitations of machine learning techniques, and Deep Learning which is a subfield of machine learning, requires a large amount of data for better results but it does not require manual feature engineering and can catch newly created phishing URLs. Deep Learning delivers what it promised when there is a big dataset to apply the Deep classifiers on.It has proven to perform well than the known traditional methods. However, its performance is still tied to a large amount of data and this single issue of data has limited the use of it because ofdata needed for and suitable to a problem a user intends to solve at a particular time.

Having seen the human contributions to the successful phishing attacks, the user education approach is introduced and used as one of the approaches needed to combat cyber threats such as phishing attacks [9] by improving users' ability to detect phishing emails and websites to overcome the attack.The email users however need all the assistance available from their email clients to avoid phishing attacks because when a phishing attack occurs, all the technical protection systems deployed cannot stop a user from disclosing personal sensitive information to the phisher over the phone or email, and of course, this attack cannot be successful without human. Therefore, much attention on human element of cyber-security is required to avoid this attack. Educating users' regularly is a good step but that would still not be enough if the same users are not tested on a regular basis which is costlierfortheorganisation.

Therefore, there is a need to adopt a holistic approach to mitigating phishing attacks because with that every security factor would be improved and secured.

## 4. CONCLUSION

The existing phishing mitigation techniques are reviewed in thispaper.Based on the literature reviewed, it can be seen as it is evident that the existing solutions have not achieved the expected decrease of phishing attacks due to the fact that the human security factors that phishers exploit often have not received an easy to use and identify phishing email. Users fall for this attack as ordinary web browsing users are not aware of how phishing attacks start or how to visually recognize illegitimate websites to differentiate them from legitimate ones[16]. The existing solutions are either residing in the serversor installed in the users' system and what the systems do are not known to the user, only the decision of the system would determine whether the user will continue or not, such as blacklist and whitelist which checks the requested URL by comparing it to what is listed in. However, with the identified downside of Blacklist, it cannot detect correctly if the URL is not listed and in such cases, the users still believe this system because the decision of the system is not visible to them. The user education technique which was introduced to help novice users to be aware of the circumstances of phishing attacks, that they may be able to minimise or avoid this risk, perhaps stop it as early as possible, also has a limitation in that, users' knowledge retention on what is taught about phishing attack and how to protect themselves from such attack. Therefore, phishing detection research should be geared towards users ease of use and identify phishing attack by developing a system that can display originality and malicious nature of both email and website. An intelligent system that can handle every factor of cyber-security should be considered. Improving the rule-based approach using Deep Learning should be considered to improve its detection.Researchers should focus on stylometric approach to reveal authors of emails that go to users 'inboxes. This paper reveals both the email and website phishing solution in phishing attack detection and provides a literature analysis of different existing phishing mitigation approaches.

# REFERENCES

[1] Leite C., Gondim J. J. C., Barreto P. S., and Alchieri E. A., (2019). Waste flooding: A phishing retaliation tool

[2] Xiujuan W., Chenxi Z., Kangfeng Z., Haoyang T., &Yuanrui T.(2019)detecting spear-phishing emails based on authentication

[3] Duman S, Kalkan-Cakmakci K, Egele M. (2016)EmailProfiler: Spear phishing filtering with header and stylometric features of emails.

[4] Calix K., Connors M., Levy D., Manzar H., McCabe G., & Westcott S. (2008). Stylometry for E-mail author identification and authentication

[5] Gupta B. B., Arachchilage N A.G., &Psannis K. E. (2018).Defending against phishing attacks: taxonomy of methods, current issues and future direction

[6] Dewan P, Kashyap A, &Kumaraguru P. (2014). Analysingsocial and stylometric features to identify spear phishing emails

[7] AbahussainO. &Harrath Y. (2019). Detection of malicious emails through regular expressions and databases

[8] Helmi R. A. A., Ren C. S.&Jamal A. (2019). Email anti-phishing detection application

[9] Asanka N. G.A.,Steve L.&Beznosov K. (2016) Phishing threat avoidance behaviour: An empirical investigation

[10] Mohammad R., Thabtah F. & McCluskey L. (2015): Tutorial and critical analysis of phishing websites methods

[11] Heartfield Ryan& George Loukas, (2018) Detecting semantic social engineering attacks with the weakest link: Implementation and empirical evaluation of a human-as-a-security-sensor framework

[12] Baniya T., Gautam D.& Kim Y. (2015). Safeguarding web surfing with URL blacklisting

[13] Canova G., Volkamer M., Bergmann C., &Borza R. (2014). NoPhish: An anti-phishing education app

[14] Bottazzi G., Casalicchio E., Marturana F., &Piu M. (2015). MP-shield: A framework for phishing detection in mobile devices.

[15] Li, J., Li, J., Chen, X., Jia, C., & Lou, W. (2015) Identity-based encryption without sourced revocation incloud computing

[16] Qabajeh I.,Thabtah F.,&Chiclana F. (2018) A recent review of conventional vs. automated cybersecurity anti-phishing techniques

[17] Lötter Andrés.&Futcher Lynn, (2015) A framework to Assist Email Users in the Identification of Phishing Attacks

[18] Gascon H., Ullrich S., Stritter B. &Rieck K. (2018) Reading between the lines: content-agnostic detection of spear-phishing emails

[19] Smadi S., Aslam N., & Zhang L. (2018). Detection of online phishing email using dynamic evolving neural network based on reinforcement learning

[20] Chandrasekaran M., Narayanan K., andUpadhayayaS. (2006) Phishing e-mail detection based on structural properties.

[21] Ghafir I., Saleem J., Hammoudeh M., Faour H., Prenosil V., Jaf S., Jabbar S. & Baker T. (2018). Security threats to critical infrastructure: the human factor

[22] Khonji M, Iraqi Y& Jones A. (2011). Mitigation of spear phishing attacks: A Content-based Authorship Identification framework

[23] Iqbal F, BinsalleehH&Fung B C M. (2010). Mining writeprints from anonymous e-mails for forensic investigation

[24] Lyon, J.& Wong M. (2006). Sender ID: authenticating e-mail," RFC 4406.

[25] KunjuM.V., Esther D., Anthony H. C. &BhelwaS. (2019) Evaluation of phishing techniques based on machine learning

[26] Peng T., Harris I., &Sawa Y. (2018).Detecting phishing attacks using natural language processing and machine learning

[27] SahingozO.K.,Buber E., Demir O., &Diri B. (2019). Machine learning based phishing detection from URLs

[28] Zhang, Y., Hong, J. I., &Cranor, L. F.(2007). Cantina: A content based approach to detecting phishing web sites.

[29] Suganya V. (2016): A review on phishing attacks and various anti-phishing techniques

[30] Abdelhamid N., Ayesh A. &Thabtah F. (2014) Phishing detection based associative classification data mining

[31] SternfeldUri&Striem-Amit Yonatan. (2019) Prevention of rendezvous generation algorithm (RGA) and domain generation algorithm (DGA) malware over exiting internet services.

[32] Akarsh S., Sriram S., &Poornachandran P.(2019) Deep learning framework for domain generation algorithms prediction using long short-term memory.

[33] Bagui S., Nandi D.,Subhash B. & White J.R (2019) Classifying phishing email using machine learning and deep learning

[34] Jain Kumar Ankit. & Gupta B.B. (2018). A machine learning based approach for phishing detection using hyperlinks information

[35] Vinayakumar R., Soman K. P., Poornachandran P., Akarsh S. &Elhoseny M. (2019)  Deep learning framework for cyber threat situational awareness based on email and url data analysis.

[36] Park Gilchan and Rayz Julia (2018).Ontological detection of phishing emails

[37] Surbhi G., Abhishek S.&Akanksha K. (2016). A literature survey on social engineering attacks: phishing attack

[38] Jamil A., Asif K.& Ghulam Z. (2018) MPMPA: A mitigation and prevention model for social engineering based phishing attacks on facebook

[39] Platsis George, (2018) Thehuman factor: Cyber security's greatest challenge

[40] NaimBaftiu. (2017).Cyber security in Kosovo

[41] Abdelhamid N., Thabtah F. & Abdel-jaber H. (2017) Phishing detection: A recent intelligent machine learning comparison based on models content and features

[42] Alsharnouby M., Alaca F., Chiasson S. (2015)Why phishing still works: User strategies for combating phishing attacks

[43] Chou N., Ledesma R., Teraguchi Y., Boneh D., and Mitchell J. C. (2004) "Client-side defence against web-based identity theft".

[44] Prakash P., Kumar M., Rao R. K. and Gupta M. (2010) PhishNet: Predictive blacklisting to detect phishing attacks

[45] Delany Mark, (2007) Domain-based email authentication using public keys advertised in the DNS (Domain Keys).

[46] Saidani N., Adi K. and AlliliM. S. (2020)A semantic-based classification approach for an enhanced spam detection.

[47] Bhowmick A. and Hazarika S.M. (2016) Machine learning for e-mail spam filtering: review techniques and trends